\documentclass[%
 reprint,
superscriptaddress,
 amsmath,amssymb,
 aps,
prb,
floatfix,
]{revtex4-1}
\usepackage{graphicx}
\usepackage{dcolumn}
\usepackage{bm}

\newcommand{\bit}{\begin{itemize}}
\newcommand{\eit}{\end{itemize}}
\newcommand{\ben}{\begin{enumerate}}
\newcommand{\een}{\end{enumerate}}

\newcommand{\be}{\begin{equation}}
\newcommand{\ee}{\end{equation}}
\newcommand{\bee}{\begin{eqnarray}}
\newcommand{\eee}{\end{eqnarray}}

\newcommand{\ba}{\begin{align}} 
\newcommand{\ea}{\end{align}}
\newcommand{\bc}{\begin{center}}                 
\newcommand{\ec}{\end{center}}
\newcommand{\bn}{\\\\$\begin{aligned}\centering}
\newcommand{\en}{\end{aligned}$}
\newcommand{\bal}{\begin{aligned}}
\newcommand{\eal}{\end{aligned}}

\newcommand{\tx}[1]{\text{#1}}

\newcommand{\ff}[2]{\frac{#1}{#2}}

\newcommand{\vv}[1]{\mathbf{#1}} 

\newcommand{\bra}[1]{\langle #1 |}
\newcommand{\ket}[1]{| #1 \rangle}

\newcommand{\exv}[1]{\langle #1 \rangle}

\newcommand{\on}[1]{\operatorname{#1}}

\renewcommand{\k}{\mathbf{k}}

\newcommand{\td}{\tx{d}}
\newcommand{\tp}{\partial}
\newcommand{\te}{\tx{e}}

\newcommand{\da}{\downarrow}
\newcommand{\ua}{\uparrow}
\newcommand{\ra}{\rightarrow}

\newcommand{\eps}{\epsilon}

\newcommand{\n}{\nu}
\newcommand{\la}{\lambda}
\newcommand{\La}{\Lambda}
\newcommand{\al}{\alpha}

\newcommand{\de}{\delta}
\newcommand{\De}{\Delta}

\begin{document}

\preprint{APS/123-QED}

\title{Heat Current Characteristics in Nanojunctions with Superconducting Leads}

\author{D. Oettinger}
 \affiliation{Institute of Mechanical Systems, ETH Z\"urich CH-8092 Z\"urich, Switzerland}
  \author{R. Chitra}
 \affiliation{Institute for Theoretical Physics, ETH Z\"urich, CH-8093 Z\"urich, Switzerland}
\author{Juliana Restrepo}%
 \affiliation{Grupo de Sistemas Complejos, Universidad Antonio Nari\~no, Medellin, Colombia}

\date{\today}

\begin{abstract}
As a fundamental requisite for thermotronics, controlling heat flow has been a longstanding quest in solid state physics.
Recently, there has been a lot of interest in nanoscale hybrid systems as possible candidates for thermal devices. 
In this context, we study the heat current in the simplest hybrid device of a two level system
weakly coupled to two heat baths. We use the reduced density matrix approach together with
a simple Born-Markov approximation to calculate the heat current in the steady state.
We consider different kinds of reservoirs and show that the nature of the reservoir plays a
 very important role in determining the thermal characteristics of the device. In particular, we investigate the effectiveness of a conventional superconductor as a reservoir with regard to manipulating the heat current. In the emergent temperature characteristics, we find that superconductivity in the reservoirs leads to
enhanced thermal currents and  that the superconducting phase transition is clearly visible in the heat current. We  observe  negative differential thermal conductance and  a pronounced rectification of the heat current, making this a good building block for a quantum thermal diode.

\end{abstract}
%
\maketitle

\section{Introduction}

The past decade has seen rapid progress in the field of engineered nanodevices. Many theoretical
proposals for ultra small quantum machines have been made, ranging from quantum heat engines\cite{Cees2005}, quantum refrigerators\cite{Fazio2014}  to
thermoelectronic devices\cite{Roberts.2011}. 
While a high degree of control of electric currents has been achieved, manipulation of heat currents is still an open problem. The ability to control heat flux would have important technological ramifications.  For example, efficient heat disposal could be built into processors, allowing us to build even smaller chips, and  construct better energy saving devices.  A thermal analog of electronics, i.e., thermotronics has also been envisaged along with thermal gates and circuits for information processing\cite{Wang.2008}. 

  
    A fundamental building block of thermotronics is the thermal diode, a rectifying device, which allows preferential flow of heat current in
one direction.\cite{Wang.2008}. 
There exist  various theoretical proposals  for  realizing efficient thermal diodes  in purely classical as well as quantum systems \cite{Roberts.2011,Terraneo.2002, Li.2004}. More recently,  the first  observations of thermal rectification in nanosystems
 \cite{Chang.2006,Scheibner.2008} followed by a realization of a quantum dot heat transistor 
were reported\cite{Saira.2007}.   
 

Progress in
the field of nanodevices depends strongly on the understanding of heat and/or charge transfer in small quantum systems
coupled to multiple thermal reservoirs.
These systems are typically 
 out of equilibrium and are no longer  described by equilibrium statistical mechanics.  This field has recently received
 a lot of attention. The presence of more than one reservoir leads
 leads to many novel phenomena, like the generation of   steady state entanglement in a two-qubit system  \cite{Sinaysky.2008,Quiroga.2007}, or novel non-equilibrium phase transitions  in one-dimensional spin chains connected to two reservoirs at their extremities\cite{Prosen.2008}.  In this paper, we focus mainly on the energy transport that can be realized in 
small quantum systems which are out of equilibrium.
 

The simplest  nanodevice capable of heat transfer can be modeled as a qubit  weakly coupled to 
two thermal reservoirs maintained at different temperatures. 
 The difference in
the temperatures results in a  steady state heat current flowing through the system. Depending on the dynamics of the qubit
and the reservoirs,  this steady state heat current can be generated either  by  simple energy exchange between the reservoirs and the
qubit,  and/or  by additional transfer or electronic charges across the junction bridging the reservoirs. 
In a series of articles\footnote{See e.\,g. \cite{Segal.2005,Wu.2009b,Wu.2009}}, Segal and coworkers used the weak coupling open system formalism\cite{Breuer.2007,Weiss.2008}, to obtain   a simple expression for the steady state heat current  passing through such a system\cite{Wu.2009}. 
 Among the analyzed examples were the generalized spin-boson model \cite{Segal.2005} and a two-level system coupled to metallic or spin baths \cite{Wu.2009}.  They found that the heat current increased with average temperature in all the systems they studied. However,  the models  studied in Ref.\onlinecite{Wu.2009} make for poor thermal diodes due to the rather weak rectification of heat current seen.   In this paper, we discuss
a model for  efficient quantum thermal diodes which involves only energy exchange.
 
 To explore the possibility of  obtaining more efficient nanodevices, a typical approach  is to replace the qubit  linking the two reservoirs  by
a more complex entity.  Here, we present an alternative approach where we use more complex reservoirs and use the properties of the reservoir  rather than the qubit(s) to obtain novel results for the heat current.
The standard boson/electron reservoirs are replaced by superconducting reservoirs
which undergo the normal metal-to-superconductor phase transition at finite temperatures. As discussed in 
Refs. \onlinecite{Camalet.2007, Restrepo.2013}, both superconductivity and phase transitions in the reservoir have
enormous impact on the dynamics of the qubit, resulting in an anomalous decay of the qubit coherence,
with associated reentrant behaviour at different temperatures in the superconducting phase.
Here, we  study  the impact of superconductivity and phase transitions on the heat current  flowing through a
qubit coupled to two superconducting  reservoirs. We find that the heat current is
extremely sensitive to superconducting order and exhibits highly non-monotonic behaviour in the vicinity of the phase
transition. This results in a fairly substantial negative differential thermal conductance.
  We also analyze the heat current when the qubit is coupled to standard metallic and insulating
reservoirs. Based on these results, we find that  a qubit coupled to one superconducting and one metallic reservoir is  a 
good model  for a  quantum  thermal diode  satisfying  multiple criteria for what constitutes a good thermal diode.

The paper is organized as follows:  In Sec. \ref{wcf}, we present the model and the  general weak coupling formalism used to study a 
qubit coupled
to two reservoirs. In Sec. \ref{sec:heatcurrentexpr}, we derive the expression for the steady state heat current and  calculate the heat current
for various reservoir setups. This is followed by a discussion of the rectification properties  of a quantum thermal
diode i.e., the qubit connected to a superconducting reservoir on the left and a metallic reservoir on the right.

\section{Setup and Weak coupling Formalism}\label{wcf}
\begin{figure}
\includegraphics[width=0.66\linewidth]{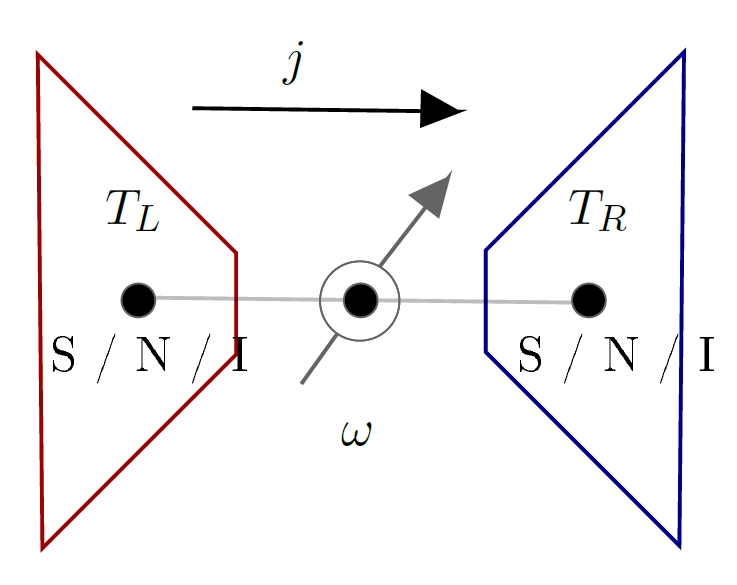}
\caption{\label{fig:setup02} Two-terminal junction consisting of two baths at temperatures $T_L$, $T_R$ and the steady state heat current $j$ transmitted via a two-level system with level splitting $\omega$. Each reservoir can be a BCS superconductor (S), a normal metal (N) or an insulator (I).}
\end{figure}

In this section, we present the formalism that describes a qubit (two-level system) weakly coupled to two  fermionic reservoirs at 
thermal equilibrium with temperatures $T_L$ and $T_R$ (cf Fig. \ref{fig:setup02}).  We use units such that the Planck and Boltzmann constants are $\hbar=k_B=1$. The total Hamiltonian describing the combined system of the qubit and two baths is given by:
\be H =  H_S+ H_B^L+  H_B^R + V^L + V^R \label{hami} \ee
The qubit is  subjected to a field in the $z$-direction and its Hamiltonian is 
\be H_S = \tfrac{1}{2}\omega \sigma_z ~. \ee 
 $H_B^{L,R}$ represent the left and right bath Hamiltonians and will be specified later. The qubit is coupled to two baths on the left and right through an Ising spin-spin  interaction
\be 
V^{L,R}\equiv S \otimes B^{L,R}=\la_{\nu}\,\sigma_x\otimes S_{x}^{L,R} ~. \label{eq:xxinteraction}
\ee
The qubit operators $\sigma_{x,z}$  are Pauli matrices  and the bath operators $S_{x}^{L,R}$ are the $x$-component of the local spin operator for the bath electrons at some origin, i.e., $S_{x}^{\nu}=\sum_{\k,\k'} ( c_{\k \da,\nu}^{\dagger}c_{\k'\ua,\nu}+c_{\k\ua,\nu}^{\dagger} c_{\k'\da,\nu} )$, with $ c^\dagger_{\k\al,\nu}$ ($c_{\k\al,\nu}$) creating (destroying) an electron of quasi-momentum $\k$ and spin $\al$ in bath $\n$. A factor of 1/2 has been absorbed into the coupling constant $\la_{\nu}$. Although we consider the case of separable coupling $S \otimes  B^{L,R}$,  the results obtained can be easily generalized to other couplings. 


In this paper, we consider electronic reservoirs which can be metallic,
insulating or superconducting. 
For  reservoirs which are either simple metals (N) or band insulators (I)  the bath Hamiltonian is given by:
\be 
H_B=\sum_{\k,\al} \eps_{\k\al}c_{\k \al}^{\dagger}c_{\k \al}~.
\ee 
 Here we drop the bath  index $\nu$ for simplicity. The electronic dispersion $\eps_{\k\al}$ then defines  the density of states (DOS)  $D(E)=\sum_{\al,\k}\de(E-\eps_{\k\al})$. The DOS for the simplest model of a metal can be taken as
\be D_N(E) =  N \left\{
     \begin{array}{lrr}
       1 &, & E \in ( -\La,\La ) \\
       0 &, & \tx{otherwise~,}
     \end{array} \right.  \label{eq:DOS:N}\ee
 with some cutoff energy $\La$. For a toy model of a band insulator with gap $2\De$, we take
\be D_I(E) = N \left\{
     \begin{array}{lrr}
       1 &, &|E| \in ( \De,\La ) \\
       0 &, & \tx{otherwise~.}
     \end{array} \right.  \label{eq:DOS:I}\ee 
where $N$ is a normalization factor. These expressions should be assumed when we refer to a metal or insulator in the following sections. 

For a reservoir exhibiting   BCS superconductivity (S) the Hamiltonian is
\be 
H_B = \sum_{\k,\alpha} E_{\k} \gamma_{\k \alpha}^{\dagger} \gamma_{\k \alpha}
\ee
with Bogoliubov quasiparticle energies  $E_{\vv k} = \on{sgn}(\xi_{\k})\sqrt{\xi_{\k}^2+\De^2}$,  where $\xi_{\k}$ is the non-interacting electron dispersion. $\De$ is the superconducting gap which is non-zero for all
$T< T_C$ where $T_C$ is the critical temperature. The fermion quasiparticle operators $\gamma$, $\gamma^{\dagger}$ are related to the electron operators via \cite{Tinkham.1996}
\bee
\gamma_{\k\alpha}^{\dagger} &=& u_{\k} c_{\k\alpha}^\dagger + v_{\k}c_{-\k-\alpha}, \\
\gamma_{-\k\alpha}&=&u_{\k}c_{-\k\alpha}-v_{\k}c_{\k -\alpha}^{\dagger},
\eee 
with $u_{\k}^2=\ff 12(1+\ff{\xi_{\k}}{E_{k}})$, $v_{\k}^2=\ff 12(1-\ff{\xi_{\k}}{E_{k}})$.
The superconducting density of states is given by 
\be D_S(E) =N\left\{
     \begin{array}{lrr}
        \ff{|E|}{\sqrt{E^2-\De^2}} &, & |E| \in ( \De ,\La ) \\
       0 &,& \tx{otherwise~}
     \end{array} \right.  .\ee
    Here, $\La$ is the Debye frequency providing a cut-off for the available energy of superconducting electrons relative to the Fermi energy. The electron density of states at the Fermi level is denoted by $N=D_N(0)$.  We see that the superconducting density of states features a gap of size $2\De$, together with a square-root singularity at $E=\pm\De$. 
For a given temperature $T=\beta^{-1}$, the gap is self-consistently determined by
\be 1 = g N \int_{0}^{\La}d E \ff{\tanh{(\beta \sqrt{E^2+\De^2}/2)}}{\sqrt{E^2+\De^2}} . \label{eq:gapequation}\ee 
with  $g$ being the strength of the  attractive coupling between electrons mediated by phonons. 
To obtain the numerical results discussed later, we choose BCS reservoirs with  a realistic value of  $gN = 0.33$ and
$T_C=0.056\La$.


The weak coupling approach is applicable to both interacting and non-interacting  bath Hamiltonians  and different couplings, provided that the energy  scale associated with the qubit-bath couplings $\lambda_{L,R}$ is smaller than all the other energy scales in the Hamiltonian.  Since we are interested in the properties of the steady state, we need to obtain the asymptotic density
matrix describing the qubit, which 
 is necessary to calculate expectation values of 
 physical observables. The  formalism described below permits one to obtain the asymptotic
density matrix  for a qubit weakly coupled to the baths.

\subsection{Born Markov master equation}\label{sec:bmme}
The total density matrix $\rho(t)$ satisfies the Liouville-von Neumann equation:
\be
 i\partial_t\rho(t)=[H,\rho(t)].
 \label{liouville}
 \ee  
The time evolution of the reduced density matrix of the qubit $\rho_S$ is obtained by taking the partial trace over both baths' degrees of freedom:
\be
\rho_S(t)=\on{Tr}_B \,[e^{i H t} \rho(t) e^{-i H t}]
\ee
For weak coupling to the baths, the reduced density matrix is found to obey a quantum master equation. In general, methods
like the  Time Convolutionless  (TCL) and the Nakajima-Zwanzig (NZ) approach\cite{Breuer.2007} can deal with both Markovian and non-Markovian dynamics. The accuracy of these schemes depends on the problem studied, making it difficult to assert a priori which one is more appropriate\cite{Breuer2004,Restrepo.2011}.  In the present problem, since the qubit has  intrinsic dynamics $H_S\neq0$, we anticipate a Markovian time evolution of the reduced density matrix at long times \cite{Restrepo.2013}.  This evolution is
well described by the usual  Born-Markov master equation derived below.  We assume that, at time $t=0$, the qubit is in a pure
state and uncorrelated to the  baths. Furthermore, the baths are in thermal equilibrium at temperatures $T_{L,R}$ respectively. The initial
density matrix is $ \rho(0) = \rho_B^L(0) \otimes \rho_S(0) \otimes \rho_B^R(0)$, where
\be
\rho_B^{L,R}(0)=\frac{e^{-H_B^{L,R} /T_{L,R}}}{\text{Tr}[e^{-H_B^{L,R} /T^{L,R}}] }
\ee
and 
\begin{align*}
\rho_S(0)=&|\alpha|^2\ket{\da}\bra{\da}+|\beta|^2\ket{\ua}\bra{\ua}+\alpha\beta^*\ket{\da}\bra{\ua}+\alpha^*\beta\ket{\ua}\bra{\da}.
\end{align*}
Here $\ket{\da}$ and $\ket{\ua}$ are the two basis states  and $\alpha$, $\beta$ are complex numbers.   Since
the reservoirs are thermodynamically large,  at weak coupling  we  use the Born approximation: This implies the bath density matrices remain almost unchanged i.e.,  $\rho_B^{L,R}(t)=\rho_B^{L,R}(0)$,
and leads to the separability of the total density matrix at long enough  times.  A final simplification is to consider an initial density matrix that commutes with the interaction, i.e.,  $Tr[V^{L,R}(t),\rho(0)]=0$. This can be achieved by renormalizing the original interaction and self-Hamiltonians. With all the above approximations, the Born-Markov master equation reads:
\begin{equation}\partial_t\rho_S(t)=-\int_0^{\infty}\sum_\nu \td s\,g_\nu(s)[S(t),S(t-s)\rho_S(t)]+h.c. 
\label{eq:master4:final}
\end{equation}
The time-dependent operators $S(t)$ and $B^\nu(t)$ are defined in the interaction picture.
Here $g_\nu(s)=\exv{B^\nu(s)B^\nu(0)}_{B^\nu}$ for $\nu=L,R$ is the two-time correlation of the bath operator $B^\nu$ and $\exv{...}_{B^\nu}=\on{Tr}_{B^\nu}(...\,\rho_B^\nu)$.

The master equation (\ref{eq:master4:final}) can be solved analytically and numerically for different time regimes. In particular, the qubit eventually loses its coherence and relaxes to its asymptotic steady state. The weak coupling formalism permits us to calculate the time
scales for decoherence and relaxation\cite{Breuer.2007}. In the Markovian regime, the attained steady state is independent
of the initial condition. In this study, since we are interested in the steady state heat current, it suffices to know the populations of the two eigenstates of $H_S$.  Eq. \ref{eq:master4:final}  then leads to the  Pauli master equation for the populations $P_{\ua}=\rho_{S\ua\ua}$, $P_{\da}=\rho_{S\da\da}=1-P_{\ua}$: 
\be 
\partial_t P_{\ua} = \sum_{\nu} (P_{\da} k_{\nu,\da\ra\ua} - P_{\ua} k_{\nu,\ua\ra\da}) . \label{eq:master:pauli}\ee
with the  rates $k_{\nu,n\ra m}$ defined  by:
\be 
k_{\nu,n\ra m}=\int_{-\infty}^{\infty}\td t\, \te^{i(E_{n}-E_m)t}\exv{B^{\nu}(t)B^{\nu}(0)}_{B^\nu}
\label{eq:rate}
\ee
These describe transitions between the qubit  states $n$, $m$ with energies $E_{n}$, $E_m$, induced by a bath $\nu$.  We denote the relaxation rate by $k_{\nu}=k_{\nu,\uparrow\ra\downarrow}$ and reexpress the excitation rate by $k_{\nu,\downarrow\ra\uparrow}=\te^{-\beta_{\nu}(E_\ua-E_\da)}k_{ \nu,\uparrow\ra\downarrow}$, where $\beta_{\nu}=1/T_{\nu}$. The steady state populations emerging from the Pauli master equation (\ref{eq:master:pauli}) are then given by
\be P_{\downarrow}=1-P_{\uparrow},~P_{\uparrow}=\ff{\sum_{\nu} \te^{-(E_\ua-E_\da)/T_{\nu}}k_{\nu} }{\sum_{\nu} (1+\te^{-(E_\ua-E_\da)/T_{\nu}})k_{\nu}}. \label{eq:master:pauli:qubitsol}\ee

\section{Steady state heat current model}\label{sec:heatcurrentexpr}
\subsection{Steady state heat current formula}
To obtain the heat current operator,  we use the approach of Ref. \onlinecite{Wu.2009b}, where the concept of a heat current via  a discretized continuity equation was introduced. When applied to our system, we find three different expressions for
the heat current operator $J_L$ ($J_R$)  modeling the heat transfer between the left (right) bath and the qubit:
   $J_L^{(1)}=-i[V^L,H_B^L]$,  $J_L^{(2)}=-i[V^L,H_S]$ and  the average $J_L^{(3)}=-\tfrac{i}{2}[H_S-H_B^L,V^L]$.  $J_L^{(3)}$ has been used in Refs. \onlinecite{Wu.2009b,Wu.2009} and $J_L^{(1)}$ in Ref. \onlinecite{Wichterich.2007}.  The physical currents  are given by
\be
j_\nu(t)=\exv{J_\nu}=Tr\left[\rho(t) J_\nu\right] .
\label{jleft}
\ee
Within the Born approximation used here, it is easy to show  that all three definitions lead to the same expectation value of the current in the steady state.  
  Here we adopt the definition $j_{\nu}\equiv\langle J_\nu^{(2)}\rangle={Tr}\{ i[H_S,V^{\nu}] \rho\}$.  In the steady state, since  $j_L=-j_R$,  we use the symmetrized heat current $j= \ff 12 (j_L -j_R)$ to write:
\be
j(t)=\frac{i}{2}Tr\left[\{[V^L,H_S]-[V^R,H_S]\}\rho(t)\right]
\label{currentop}
\ee


Evaluating Eq. (\ref{currentop}), we obtain the following expression for the steady state heat current:
\[ j = {\ff  \omega 2} \left[ P_{\da} (k_{L, \da\ra\ua}-k_{R, \da\ra\ua})- P_{\ua} (k_{L, \ua\ra\da}-k_{R, \ua\ra\da}) \right] \label{eq:current:steadystate}. \]
Using the solution (\ref{eq:master:pauli:qubitsol}) for the populations, the steady state current takes the compact form \cite{Wu.2009}

\be
 j(\omega,T_L,T_R)= \ff{\omega (n_L(\omega)-n_R(\omega))}{\tilde n_L(\omega)/k_{L}(\omega, T_L)+\tilde n_R(\omega)/k_{R}(\omega,T_R)}, \label{eq:jsteady:qubit} 
 \ee
where $n_{\nu}(\omega)\equiv n_{\nu}(\omega,T_{\nu})=[\te^{\omega/T_{\nu}}+1]^{-1}$ and $\tilde n_\nu(\omega)\equiv n_\nu(-\omega)$. The relaxation rate induced by the reservoir labeled by $\nu$ is $k_\nu\equiv k_{\nu,\uparrow\ra\downarrow}= k_{\nu}(\omega, T_{\nu})$.  Note  that details  of the reservoir  manifest themselves directly through the relaxation rate $k_\nu$ which depends on two-time correlation functions. Since the expression for
 the current captures essentially the sequential tunneling contribution,  the qubit level splitting  should satisfy $\omega\ll T_{\nu}$.
 In the opposite limit, where $\omega \gg T_\nu$, one is in the cotunneling regime and  the heat current can be
 obtained via the  Born-Oppenheimer approximation\cite{Wu.2011}. One expects a smooth extrapolation between these
 limits at least for weak dissipation.\\
 
\subsection{Relaxation rates for the different reservoirs}
In this subsection we study the relaxation rates $k_{\nu}(\omega,T)$ for the metallic (N), insulating (I) and superconducting (S) reservoirs. 
 For the interaction Hamiltonian considered here,  the rates can be evaluated in a straightforward manner:
 \begin{widetext}
\begin{eqnarray}
k_{\nu}(\omega, T_\nu) &= &\la_{\nu}^2\int_{-\infty}^{\infty}\td t\, \te^{i\omega t }\exv{S_{x}^{\nu}(t)S_{x}^{\nu}(0)}_{\nu} \nonumber \\
  &= &\pi \la_\nu^2\int_{-\infty}^{\infty}\td E\,f_\nu(E) n_\nu(E)[1-n_\nu(E+\omega)] D_\nu(E)D_\nu(E+\omega)   \label{eq:knu_metal}
  \end{eqnarray}
  \end{widetext}
where $f_\nu(E)=1+\Delta_\nu^2/E(E+\omega)$  for a superconducting reservoir  and $f_\nu(E)=1$ for metallic and  insulating reservoirs. $n_\nu(E)$ is the Fermi occupation number and $D_\nu(E)$ is the associated DOS. 

 In what follows, we suppress the index $\nu$  for  notational clarity.
   For  the metallic DOS  in (\ref{eq:DOS:N}), $k$ is easily found to be 
   \be 
   k(\omega,T)= 2\pi\la^2 T \ff{\te^{\omega/T}}{\te^{\omega/T}-1}\log{\Big[\cosh{\left(\tfrac{\La}{2T}\right)}\on{sech}\left(\tfrac{\La-\omega}{2T}\right)\Big]}~.
    \label{eq:knu:N}
   \ee
   
For the insulator \cite{Restrepo.2011} at zero temperature, $k(\omega,0)=0$ for $\omega \leq 2\Delta$ and $k(\omega,0) \propto\lambda^2 (\omega-2\Delta)^2$  for $\omega$ close to $2\Delta$. As temperature increases, the thermal activation of the gap results in a rate which is no longer gapped. For $\omega \ll 2\Delta$, 
\be
   k(\omega,T)= 2\pi\la^2 T \ff{\te^{\omega/T}}{\te^{\omega/T}-1}\log{\Big[\frac{\cosh{\left(\tfrac{\La}{2T}\right)}}{\cosh\left(\tfrac{\La-\omega}{2T}\right)}\frac{\cosh{\left(\tfrac{\Delta}{2T}\right)}}{\cosh\left(\tfrac{\Delta+\omega}{2T}\right)}\Big]}~.
\ee
For larger values of $\omega$ the structure of the insulating rate is very similar to that of  the metal (\ref{eq:knu:N}).


The rate for the  BCS reservoir  cannot be evaluated analytically for all frequencies $\omega$ and temperatures $0 < T < T_C$\cite{Restrepo.2013}. In particular, it is singular for $\omega \ra 0$ due to the singularity of the DOS at $E=\De$, and it shows a pseudogap due to the gap in the density of states.    For $0<\omega \ll \De \ll \La$\cite{Restrepo.2013}, (this excludes regions close to $T_C$, where the gap
$\De$ is very small),
\be 
k(\omega,T)\approx-\pi \la^2\ff{\De}{2 \cosh^2(\De/2T)} \log (\omega/T)~.
 \label{eq:kBCS:approx}
 \ee 
 For general parameter values, the rate can be obtained  numerically.


\begin{figure}[htbp]
\centering
\includegraphics[width=0.80\linewidth]{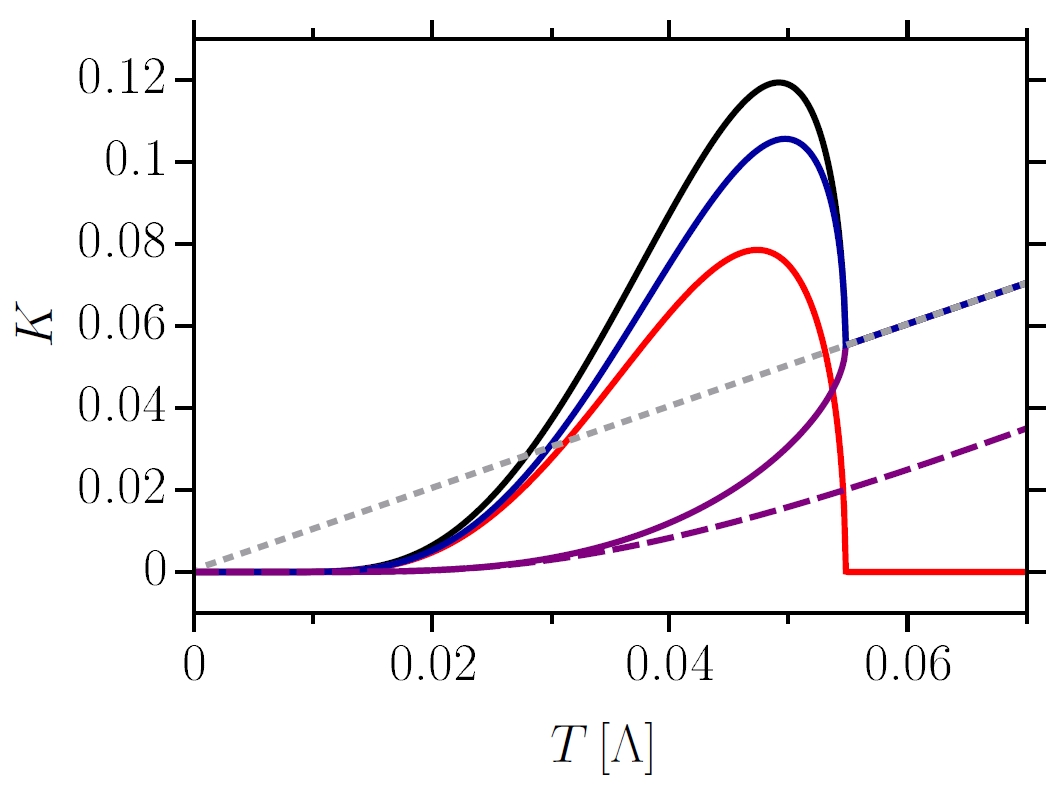}
\caption{The rate $K$ as a function of bath temperature $T=1/\beta$ for $\omega=10^{-3}\La$ for different reservoirs. Solid black:  exact BCS result from  the full integral in Eq.\,(\ref{eq:knu_metal}), red: singular analytical contribution to the BCS result from Eq.\,(\ref{eq:kBCS:approx}), dashed grey: metallic reservoir, dashed violet: insulating reservoir with fixed gap $\De = 0.095\La$, violet: insulator with  the BCS like $T$-dependent gap and  blue: $T$-dependent insulator plus singular analytical BCS contribution (\ref{eq:kBCS:approx}). }
\label{fig:TK_BCS_omega0p001}
\end{figure}
In Fig.\,\ref{fig:TK_BCS_omega0p001}, we plot  the temperature dependence of $K=k/\pi\la^2$ at fixed $\omega$ for all the aforementioned reservoir types. The first thing we note is that the BCS superconductor has a much higher relaxation rate
than an analogous metal in the entire temperature range $T < T_C$. This is attributed
to the  singularity in the density of states $D_S(E)$.  For the superconductor, we see that   (\ref{eq:kBCS:approx})  agrees with the exact numerical curve  for $T<0.04\La$. At higher temperatures, the contribution from thermal activation across the shrinking gap, neglected in  (\ref{eq:kBCS:approx}), becomes important. The latter contribution  can be well described by the  rate $K$ of  an insulator with $T$-dependent gap $\De(T)$.  Adding this to the approximation of the singular contribution (\ref{eq:kBCS:approx}), we see reasonable agreement with the exact BCS curve in the entire temperature range.

\section{Heat current results}\label{sec:curvesh}
\subsection{Heat current characteristics}\label{sec:curves}
The results for the rates presented in the previous section can be used in conjunction with Eq.(\ref{eq:jsteady:qubit}) to
obtained the heat current for various setups.
In the rest of the paper, except in Sec. \ref{subrecti}, we consider only
 the case of  symmetric couplings $ \la = \la_L =  \la_R $.\\
We now present our results for the normalized steady state heat current $j\equiv j/\pi\omega\la^2$.
  as a function of the temperature bias $\Delta T\equiv T_L-T_R$  and
 the average temperature $T_a\equiv \tfrac{1}{2}(T_L+T_R)$. 
    For each reservoir we select a normal metal (N), a BCS superconductor (S) or an insulator (I) with a constant or a T-dependent gap.  The setup (Fig. \ref{fig:setup02}) is identified using the abbreviations for the reservoirs: `SN' for instance means that we have a superconducting bath on the left, and a metallic bath on the right.  
 For most plots we only consider $\De T \geq 0$, i.\,e. $T_L \geq T_R$, which results in  a positive $j$.

\begin{figure}[htbp]
\centering
\includegraphics[width=0.99\linewidth]{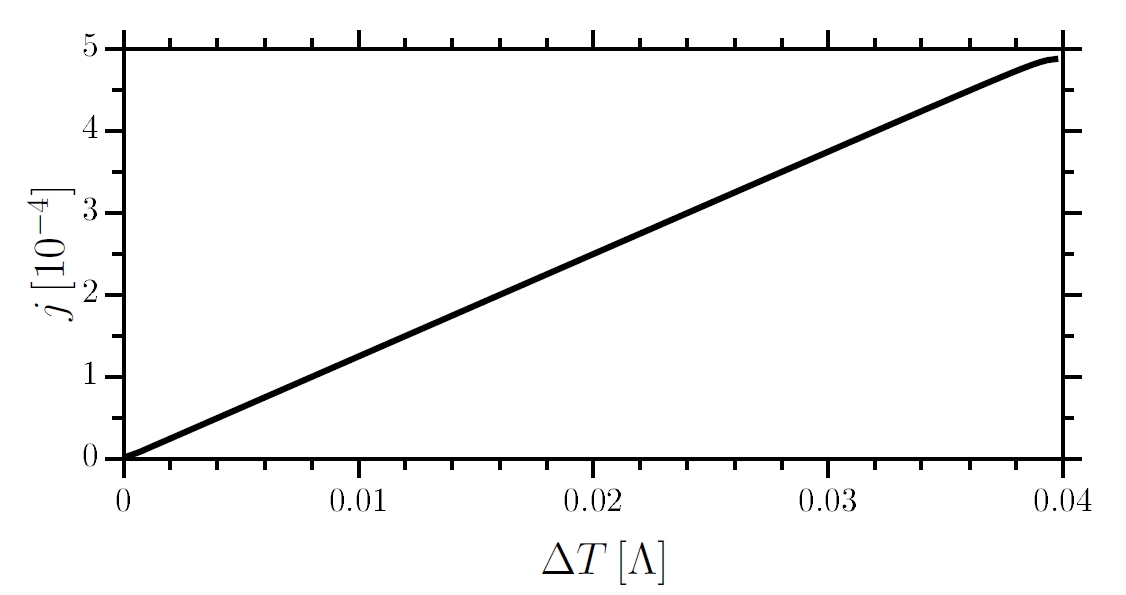}
\caption{ Heat current $j(\De T)$ for $T_a=0.02\La$, $\omega=10^{-3}\La$ for a metal-metal (NN) setup. The curve has been obtained using the analytical expression (\ref{eq:knu:N}) for the $k_{\nu}$.}
\label{fig:JDeT_NN_Ta0p02}
\end{figure}

We first show the heat current in a system with simple metallic
reservoirs (NN) in Fig.\,\ref{fig:JDeT_NN_Ta0p02}. Note that  the heat current  simply increases linearly (i.e. monotonously) with increasing temperature difference $\De T$. This is typical for most simple reservoirs whose
excitations are gapless. Similar behaviour was seen with bosonic reservoirs in Ref. \onlinecite{Wu.2009}.  


To obtain heat currents
with  non-trivial characteristics, we now 
consider superconducting  and insulating reservoirs.
 We find that there are three relevant domains for the average temperature which produce distinct heat current characteristics. In each case we discuss how a gap and/or a singularity in the DOS makes the heat current deviate from what is seen in more trivial setups like the one shown in Fig. \ref{fig:JDeT_NN_Ta0p02} for two metallic reservoirs. 
 In all three cases, the current obeys Fourier's law, $j \propto \De T$ as $\De T \to 0$, for all reservoir types. The regime of its validity however depends on the nature of the reservoirs.


\begin{figure}[htbp]
\centering
\includegraphics[width=0.99\linewidth]{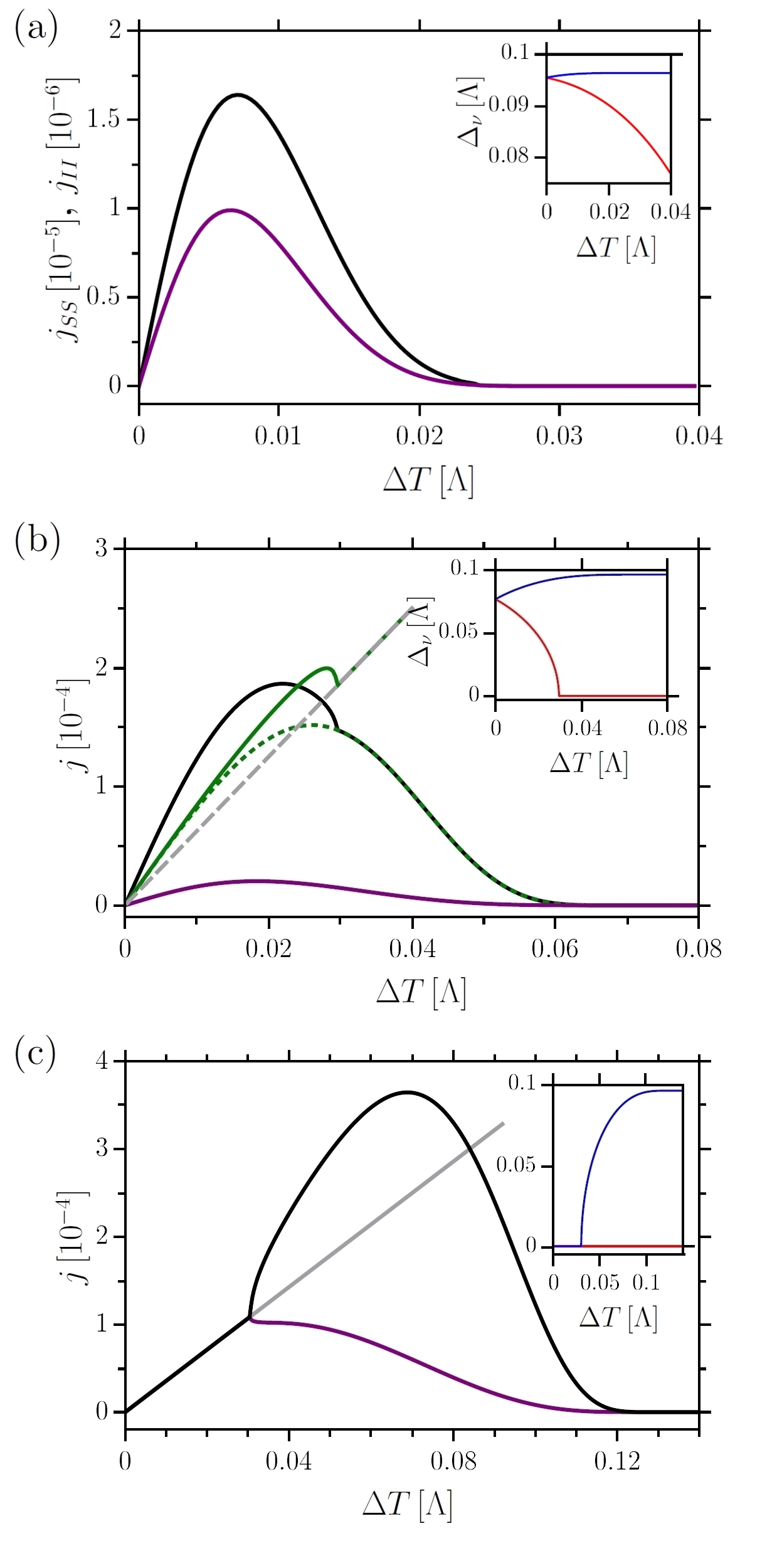}
\caption{Heat current $j(\De T)$ at $\omega=10^{-3}\La$ for various average temperatures and reservoir types. The insets show the temperature dependence of the gaps $\De_{\nu}[\La]$ in the $L$, $R$ baths (red, blue) setups involving superconducting baths or insulators with a  temperature dependent BCS like gap. \emph{Top:} $T_a=0.02\La$, black: SS setup, violet: II setup with fixed gaps $\De_L=\De_R=0.095\La$. 
\emph{Middle:} $T_a=0.04\La$, black: SS, green: SN, dashed green: NS, dashed grey: NN, violet: II setup with variable gaps as for the superconductor.
\emph{Bottom:} $T_a=0.07\La$, black: SS$=$NS, violet: NI$=$II with variable gaps as for the superconductor and  grey: NN. 
}
\label{fig:JDeT_various}
\end{figure}

\emph{Low temperatures, $T_a \lesssim T_C/2$:} In Fig.\,\ref{fig:JDeT_various}\,(a), we consider an SS setup with both
reservoirs being identical and having the same critical temperature $T_C$. We choose $T_a=0.02\La$ with $\De T\in[0,0.04\La]$, which means having two superconducting materials deep in the ordered phase for the range of possible $\De T$.
  The corresponding gaps for the two reservoirs are also plotted in the
inset. The current (black curve) in this regime is rather small due to the presence of the gaps in both reservoirs. 
   Note that in this low $T_a$ regime, the heat current for the SS setup 
is very similar to that seen in a setup with simple insulating baths (II) with a 
temperature-independent gap  $\De_{\nu}=0.095\La$ (violet curve). Both start off linearly, 
then reach their maximum values at $\De T\approx0.007\La$, and finally decay exponentially to zero as $T_R \to 0$. 
This strong suppression of the heat current is due to the increase in the size of the gap as $T_R \to 0$, which blocks
conduction more efficiently.
We conclude that for baths well in the superconducting regime, the heat current is primarily  determined by the existence of the gap and associated thermal activation. The weak temperature dependence of the gap as well as the singularity of the DOS---which are not present in the II system---have only minor effects on the qualitative features.  

\emph{Intermediate temperatures, $T_a \lesssim T_C$:} We consider
an average temperature sufficiently close to $T_C$, so that the left reservoir undergoes a superconductor-to-normal
metal transition as $\De T$ increases. This is the most interesting regime where the heat current displays highly
nontrivial behaviour. In Fig.\,\ref{fig:JDeT_various}\,(b), we consider $T_a=0.04\La$.  The evolution of
the gaps in both superconducting reservoirs is plotted in the inset. For this SS setup,  the heat current
(black curve)  increases, reaches a maximum, and then starts to decrease. As $T_L \to T_C$, the
left reservoir undergoes a transition to a normal metal which becomes visible through a pronounced
kink in $j$.  Even though the left reservoir is now metallic, the current still decreases because of the increasing gap in
the right reservoir. The appearance of a kink is a direct consequence of the phase transition in the bath.
The position of this kink can be changed by varying $T_a$.   


To understand the impact of superconductivity in more detail, we also plot the heat current for the SN and NS systems. For the SN system (green curve), we see that the imminent loss of superconductivity in the left reservoir is signaled by a downturn in the current,
followed by a pronounced kink at $T_L=T_C$.  Beyond that, we see linear behaviour of $j$ as expected for a system composed
of two simple metallic baths. The current in the NS system is however different as the superconducting gap in the right
reservoir dominates the  heat conduction and we see the behaviour reminiscent of insulating baths.


 Comparing  these
results with the heat current for insulating baths with
a $T$-dependent gap of the BCS form (cf. violet curve in Fig. 4b) , we see that  the singularities associated with the
superconducting reservoir are essential to obtain the kinks in the heat current.  
As expected, for sufficiently  high $\De T$, the results for the SS and NS setups coincide.
Having at least one superconducting bath implies an appreciably amplified heat current in a considerable range of $\De T$ even in comparison to the case of  metallic reservoirs. This can  be traced back to  the  $\on{log}$-type singularity in  the BCS rate $k_{\nu}$  which results in a rate higher than that for metallic baths (cf. Fig.\,\ref{fig:TK_BCS_omega0p001}).  For the parameters
chosen here, we  see that the heat current in the SS case  (black curve in Fig.\,\ref{fig:JDeT_various}\,(a)) is higher by a factor of $10$ as compared to insulators with a comparable gap size (violet), and remains sizably larger than in a metal-metal system (grey dashed) which in itself could already be considered a good heat conductor. We also notice that the downturn of $j$ in $\De T$ near the phase transistion is very strong for the superconducting reservoirs, see Sec.\,\ref{subrecti} for more details.

\emph{High temperatures, $T_a \gtrsim T_C$:} We first consider the NS/SS setup with both reservoirs starting above their critical temperature, at $T_a=0.07\La$. The left bath is heated, remaining metallic, and the right bath is cooled and undergoes a phase transition to
the superconducting phase---at which point we see a kink in the corresponding black curve in Fig.\,\ref{fig:JDeT_various}\,(c). A similar kink is seen in the NI setup (violet) with a temperature-dependent gap appearing for the insulator below $T_C$. This shows that the emergence of the kink in this regime is only due to the appearance of a gap in the spectrum and has nothing to do with the divergent behaviour of the rates. Going back to Fig.\,\ref{fig:JDeT_various}\,(c), we see that once the right bath reaches sufficiently low temperatures, both superconductor and insulator suppress heat flow again.
Regarding the SS setup, we observe again that for a wide range of temperatures where $T_R<T_C$, the heat current for the NS system (black) is strongly amplified compared to a setup of two metallic leads (grey curve).
 

\begin{figure}[htbp]
\centering
\includegraphics[width=0.70\linewidth]{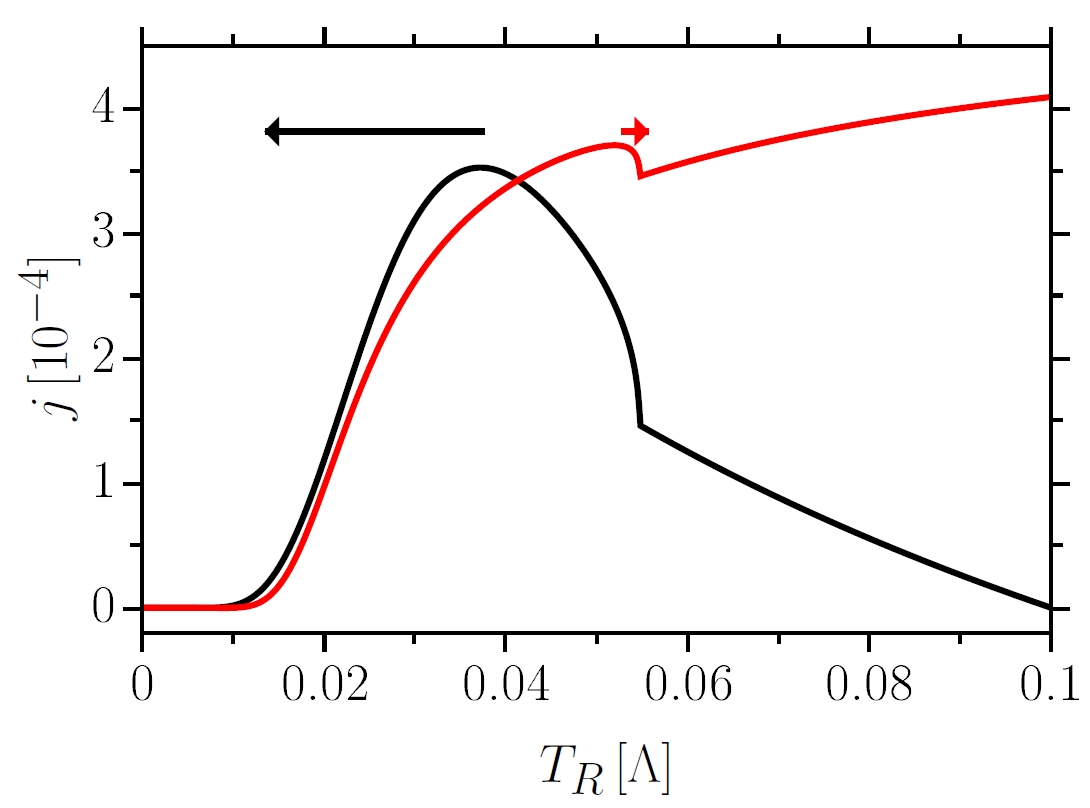}
\caption{Heat current $j$ for an NS setup with $T_L=0.1\La$ (black), $T_L=0.01\La$ (red) and variable $T_R$, $\omega=10^{-3}\La$. The regions exhibiting NDTC are indicated by arrows which also mark the direction of increasing temperature difference.}
\label{fig:j_TR_TL0p1TL0p01_w0p001}
\end{figure}
 A direct consequence of the non-monotonic nature of the heat current when a qubit is coupled to at least one
 superconducting reservoir is the occurrence of negative differential thermal resistance (NDTR) or negative differential thermal conductance (NDTC) \cite{Wu.2009, Li.2006},
  where increasing the temperature difference $\De T$ rather counterintuitively results in a reduced heat current.  
  In the problem considered, NDTC is more pronounced in the vicinity of the phase  transistion. If one considers the heat current to be a function of one reservoir temperature, say $T_R$, keeping the other one fixed \cite{Li.2006,Hu.2011}, one can quantify the NDTC by the derivative $\ff{\tp j}{\tp T_R}$. In Fig.\,\ref{fig:j_TR_TL0p1TL0p01_w0p001} we plot the  NDTC of a NS setup as a function of $T_R$ for fixed $T_L=0.1\La$ (black) or $T_L=0.01\La$ (red). In the first case, there is a large region of NDTC when the right reservoir is in the superconducting phase. The slope is however moderate compared to the second case where we undergo the phase transition while heating up. There we see a sharp downturn within a very small region of $T_R$ immediately before the phase transistion, and therefore very high NDTC. Strong NDTC has been identified as a requisite for building thermal transistors as it  determines the   amplification function of such  devices\cite{Li.2006}.  Consequently, the strong NDTC seen in
  our model in the vicinity of the superconducting phase transition potentially  makes our system a good candidate for thermal
  transistors provided it has strong rectifiying properties. The rectification of the heat current is studied in the following section.
 

\subsection{Rectification of heat current}\label{subrecti}
Encouraged by the remarkable features of the heat current characteristics for two-terminal setups with at least one superconducting bath, we now investigate how well such a system is suited to form the basic building block of a thermal circuit, the thermal diode/rectifier. Rectification can be quantified by:
\be R(\De T)= \ff{|j(\De T)|-|j(-\De T)|}{|j(\De T)|+|j(-\De T)|}, \label{eq:def:recti}\ee
where $j$ is the normalized heat current. Some authors use slightly different definitions \cite{Wu.2009,Ruokola.2009,Kuo.2010}.
 In our convention, $R=0$ means there is no rectification, $|R|=1$ corresponds to the
ideal  case where transport of thermal energy is allowed in the forward direction, and fully blocked in the reverse direction.  Several proposals for thermal rectifiers have been made for a variety of nanosystems \cite{Wu.2009,Ruokola.2009} including insulator-quantum dot-vacuum tunnel junctions \cite{Kuo.2010} and in graphene nanoribbons \cite{Hu.2011}.  Rectification of the
heat current   in quantum dot systems was also experimentally observed  \cite{Scheibner.2008}.  However,  most of these proposals 
lead to relatively low  
rectification ratios \cite{Roberts.2011},  and achieving high rectification ratios remains an open problem. 

eq:jsteady:qubit
From the expression for the
normalized heat current $j$  and  $R$  in Eqs. \ref{eq:jsteady:qubit} and \ref{eq:def:recti}, we see that 
rectification requires  the inequality
 \begin{equation}
 \left(\ff{\tilde n_L}{k_L(T_L)}-\ff{\tilde n_L}{k_R(T_L)}\right) \neq \left(\ff{ \tilde n_R}{k_L(T_R)}-\ff{ \tilde n_R}{k_R(T_R)}\right)
 \end{equation}
  to hold for general $T_L$, $T_R$. Therefore, for rectification, we need  $k_L\neq k_R$. This can be achieved via i) unequal couplings $\la_L\neq\la_R$ and identical baths, ii) different baths and equal couplings, or iii) a combination of both possible asymmetries. To obtain  a strong thermal
  rectifier in the weak coupling limit, we consider here the  second option. 
  
 
  An important first question is whether NDTC is a necessary condition for rectification: For two identical superconducting baths coupled  equally to a qubit at a given average temperature, one would have NDTC (cf. Fig.\,\ref{fig:JDeT_various}\,(a)) for
  both directions of thermal bias, but no rectification. On the other hand, we can have rectification in a system with
  asymmetric couplings to two identical  reservoirs even in the absence of any NDTC.
    However, the onset of  rectification is only linear, and thus one needs an impractically high temperature bias to get a good rectification ratio. This is due to the absence of NDTC and the resulting insufficient suppression of heat current in the one direction. 
  NDTC is however necessary for   rectification in
 the case of  symmetrically coupled non-identical reservoirs.
 
       
Based on the heat current characteristics presented in the
  previous section, a good candidate for the  thermal diode   would be  the qubit connected to a superconductor and a normal metal (SN).  Such a setup has the added advantage of higher values of current. One can also investigate the impact of the 
  superconducting phase transition on  the rectification of  the
  heat current. 
  We evaluate the rectification properties for this device at different average temperatures $T_a$ in the three
temperature domains discussed earlier.
\begin{figure}
\centering
\includegraphics[width=0.99\linewidth]{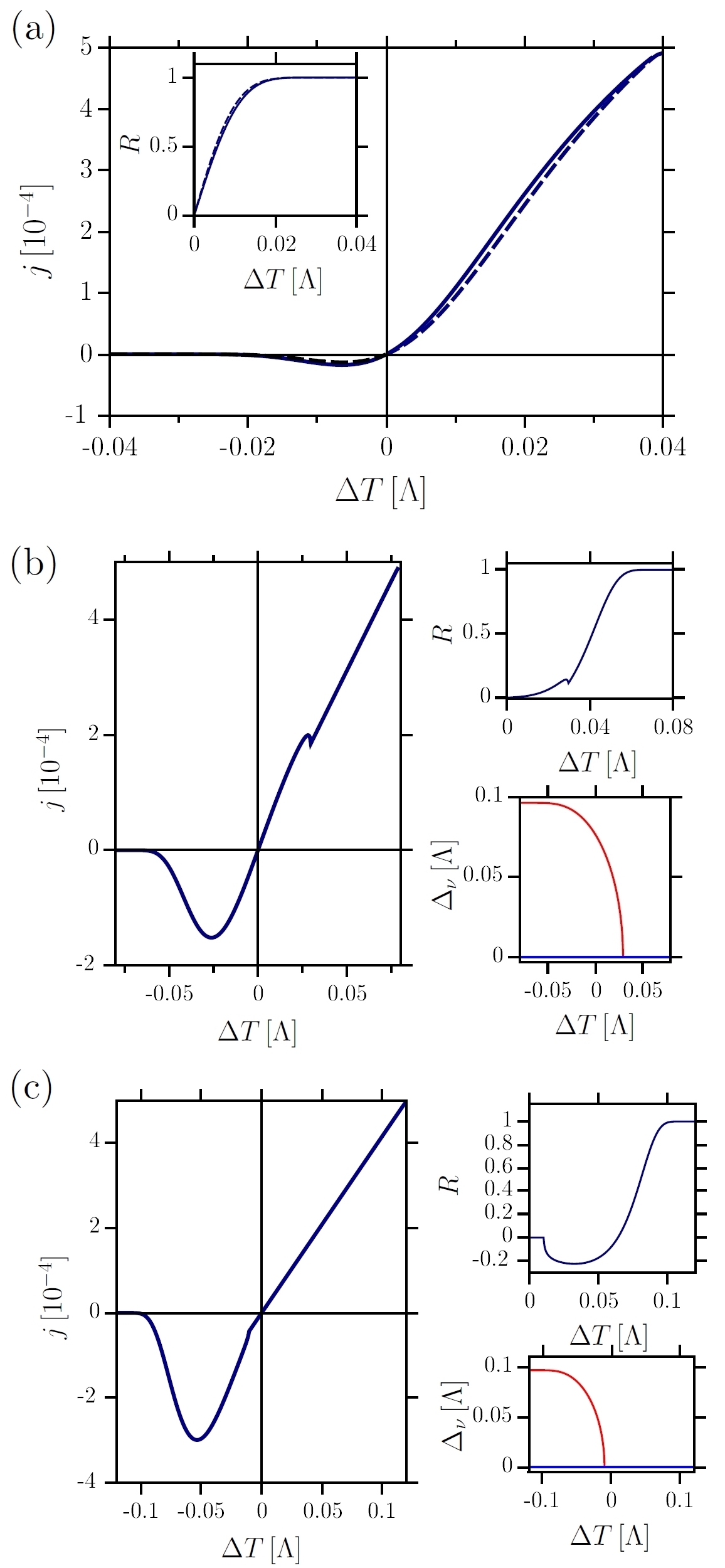}
\caption{Heat current $j(\De T)$ at $\omega=10^{-3}\La$ for the SN setup and various average temperatures. For all plots we show the corresponding rectification $R(\De T)$. For (b) and (c), we show the temperature dependence of the gaps $\De_{\nu}[\La]$ in the $L$, $R$ baths (red, blue). \emph{Top:} $T_a=0.02\La$, the solid lines represent numerical, the dashed lines analytical results. The latter are obtained using (\ref{eq:kBCS:approx}) and  (\ref{eq:knu:N}).
\emph{Middle:} $T_a=0.04\La$;
\emph{Bottom:} $T_a=0.07\La$.
}
\label{fig:Recti_various}
\end{figure}


\emph{Low temperatures, $T_a \lesssim T_C/2$:}  We again choose $T_a=0.02\Lambda$, so that the superconductor is in its
ordered phase at $\De T=0$. The heat current for both positive and negative temperature differences is plotted
in Fig.\,\ref{fig:Recti_various}\,(a).   For $\De T<0$ we see a strong suppression of heat current, rendering this the reverse direction for heat conduction. The explanation is  the same as for Fig.\,\ref{fig:JDeT_various}\,(a) where we considered the SS setup. For $\De T<0$,
the superconductor on the left is cooled towards $T_L\ra 0$ and possible transitions across the gap are  frozen, making  $k_L$  and hence $j$ very small.  For $\De T>0$,  i.e., the forward direction, we see a substantial heat current $j$, which starting from $\De T\gtrsim 0.008\La$ becomes even larger than the one we would obtain if we considered an NN setup. This can be concluded from Fig.\,\ref{fig:TK_BCS_omega0p001}, where the BCS relaxation rate becomes larger than the one for the metal from around $T\approx0.028\La$. Consequently, $R$  increases rapidly as a function of the bias $\De T$, as shown in the inset.  Moreover,  $R\sim 1$ in a considerable range of temperatures.  Comparing our results to Ref. \onlinecite{Wu.2009}
 where a maximal $R\approx 0.18$ for $\De T/T_a = 1/5$ was obtained for metallic reservoirs, our SN diode achieves a
 rectification of $R\approx 0.69$  for the corresponding thermal bias $\De T=0.008\La$. Doubling the bias  results in $R\approx 0.96$ which is nearly ideal.  This illustrates the efficiency of using reservoir properties to increase rectification.

\emph{Intermediate temperatures, $T_a \lesssim T_C$:} In Fig.\,\ref{fig:Recti_various}\,(b), we see that in the intermediate
temperature range, starting from $\De T=0$, $j$ increases almost uniformly, showing the phase transition with the known kink, and always being slightly higher or equal to the values for an NN setup.  
For $\De T<0$, the reverse direction, at first there is still a sizable heat current in a large region of $\De T$, because starting from $T_L=0.04\La$ (cf. Fig. \ref{fig:TK_BCS_omega0p001}), the underlying relaxation rates are large due to the singularity in the superconducting DOS. The current is suppressed only at   very low temperatures in the left bath.
In comparison to the low-$T_a$ case,  the large values of $j$ for a range of negative $\De T$ imply that $R$ (inset) grows more slowly with increasing bias. Because of the eventual suppression of $j$ for high negative bias, an ideal $R\approx 1$ can nevertheless still be reached. Finally note that rectification $R$ also exhibits the phase transition via a local downturn at $\De T\approx 0.03\La$.


\emph{High temperatures, $T_a \gtrsim T_C$:} Results for this regime are shown in Fig.\,\ref{fig:Recti_various}\,(c). Here, for $\De T>0$, both materials are metallic, whereas for  $\De T<0$  the left bath undergoes a phase transition to the ordered phase.
The heat current is higher in the regime where the left bath is superconducting  as compared to the $\De T > 0$ regime where
one has the usual metallic behaviour. Consequently, the rectification $R$ is negative for a wide range of $\De T$  and then
$R\approx 1$ for high $\De T$.  While in the range of negative rectification, $R$ is not ideal, it is still sizable with $R\approx -0.2$. We emphasize that this flip of direction is completely independent of the device design and solely relies on changing the reservoir temperatures. Similar reversals of the rectification have been   seen in    nonlinear circuits with an anharmonic central mode inductively coupled to two reservoirs \cite{Ruokola.2009}. However, the rectification obtained in these systems remains very weak as compared
to our SN diode.

Furthermore, combining both asymmetries: non-identical reservoirs and asymmetric couplings, we can further improve the rectification ratio of
the SN setup. To illustrate this, we 
 consider a stronger coupling of the metallic reservoir  to the qubit, i.e., $\la_R=3\la_L$. At $T_a=0.02\La$ we find a stronger
 suppression of the  current in the reverse direction as compared to the case of symmetric coupling, leading to higher rectification. For instance, at $\De T= 0.008\La$ we now have $R\approx 0.77$ instead of $R\approx 0.69$ for symmetric coupling. For intermediate temperatures, $T_a=0.04 \La$,  we again find a substantial increase of $R$.
 To summarize,  the combined asymmetry  increases rectification for all $T_a<T_C$, thus effectively extending the range of temperature where the diode works reasonably well. For $T_a>T_C$ we find that the rectification in the region of $R<1$ is reduced. Inverting the couplings to $3\lambda_L = \lambda_R$ helps increasing the negative rectification.
 Based on the results shown above, we  see that even for moderate temperature bias, the SN-device produces  reasonably high rectification and  we identify low temperatures as the preferred working region for this setup.

\section{Summary and outlook}

We have examined the heat transfer across a two-terminal junction transmitting energy through a qubit.  Using the
weak coupling formalism, we have studied the impact of different reservoirs on the
heat current through a qubit.
We find that although in the weak coupling limit, the steady state of the qubit  and the associated populations are insensitive to
the details of the reservoirs, physical observables like the steady state heat current are determined by detailed
properties of the reservoirs.  By studying metallic, insulating and superconducting
reservoirs,  we show that the heat current is in fact a good
probe of the  reservoir physics, especially  the superconducting  phase transition. Phase
transitions in the reservoir manifest themselves as 
 a kink in the heat current with an accompanying amplification of the current and a temperature regime with sizable NDTC.  
 Giving a more general perspective to the above discussion,  we note that  for any system undergoing
a continuous phase transition, there is an accompanying  change in the nature of its excitation spectrum and hence the underlying
density of states. Consequently,  we expect that the heat current will signal  transitions  in the bath by a  change of slope or curvature.
Details of this change, however, as well as the potential amplification of the heat current in the ordered phase will depend on the details of the system under study.

From a device perspective,
   our results show that the SN setup is a good candidate for an efficient
  quantum thermal diode.  This SN diode satisfies the fundamental characteristics required for a diode:

\begin{itemize}
	\item High rectification.
        \item  Higher heat currents as opposed to diodes made using metallic reservoirs.
	\item  Large NDTC, making it a good building block for a thermal transistor.
\item Short switching times  between forward and reverse bias. The time to achieve steady state following a reversal of
temperature bias is   determined by the relaxation time of the setup. Here, the BCS reservoir brings another advantage which can be argued as follows: The Markovian relaxation is described by \cite{Restrepo.2013, Juliana.2011}
$
\ln \exv{\sigma_z(t)} \propto -\sum_{\nu=L,R} \gamma_{\nu} t
$
with the relaxation rate
$
\gamma_{\nu}(\omega) \propto [k_{\nu}(\omega)+k_{\nu}(-\omega)].
$
For small splitting of the qubit levels, the BCS rate dominates because of  its singular behaviour at low fields.
Taking for instance $\omega=10^{-3}\La$, we see from Fig.\,\ref{fig:TK_BCS_omega0p001} that in a range of temperatures $0.028 \La< T< T_C$, having a BCS reservoir will allow faster relaxation and thus faster switching times compared to the case where one has  metallic baths.
	\end{itemize}
To summarize, with the goal of tailoring certain characteristics in nanodevices, it appears useful to engineer reservoirs rather than the central
system. This also raises the question of what heat current characteristics one would
expect for these nontrivial setups if one goes beyond weak coupling---which is however a very hard task. 
As directions for further work, we also suggest extensions of this study to thermoelectric devices, where one would need to look at both heat and electric currents, and to explore the physics of a thermal transistor involving three qubits coupled to
three different reservoirs.

%
\end{document}